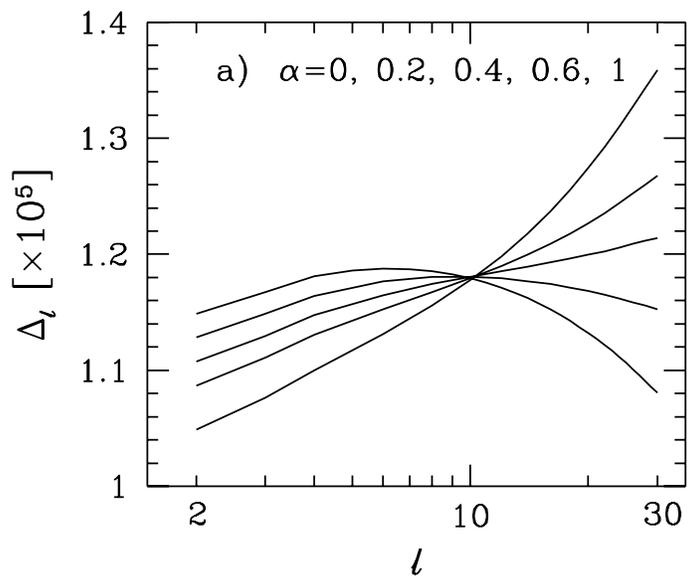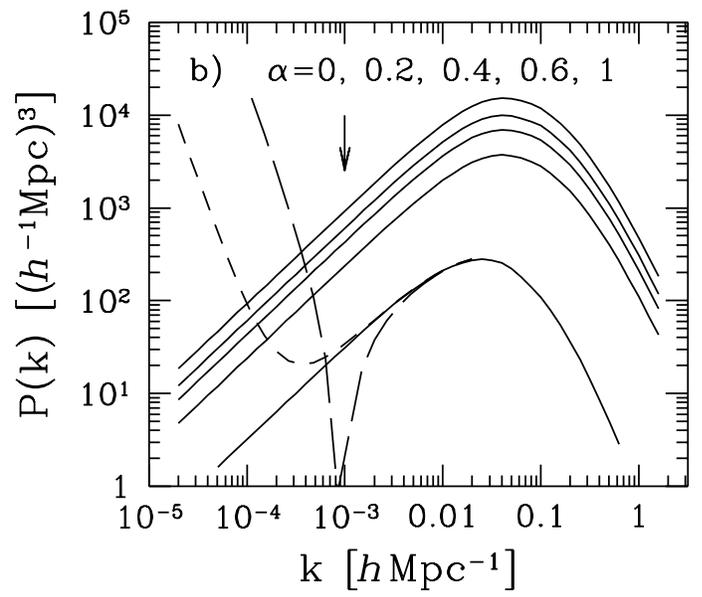

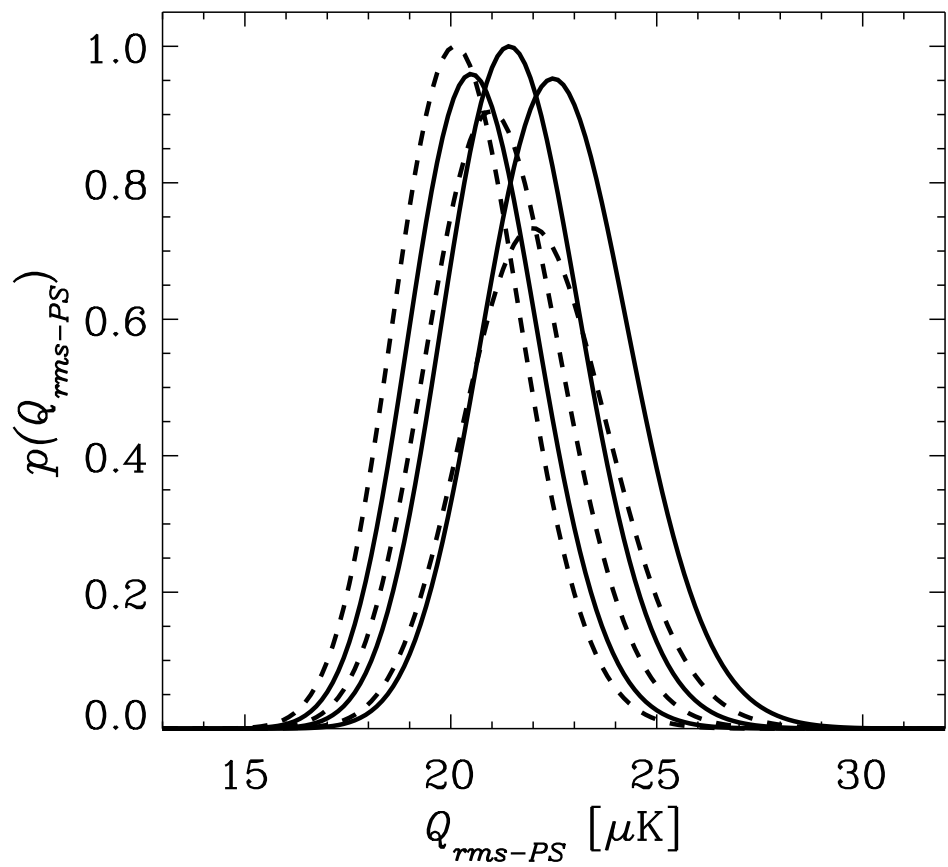

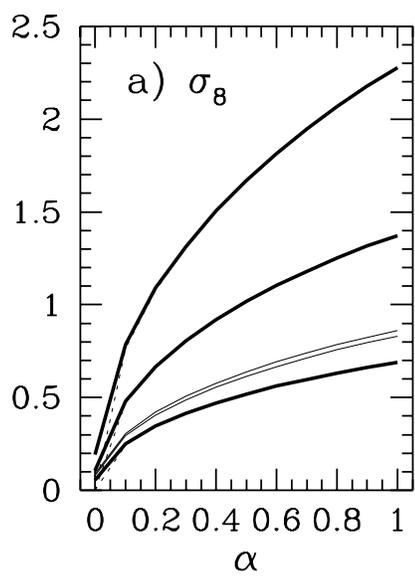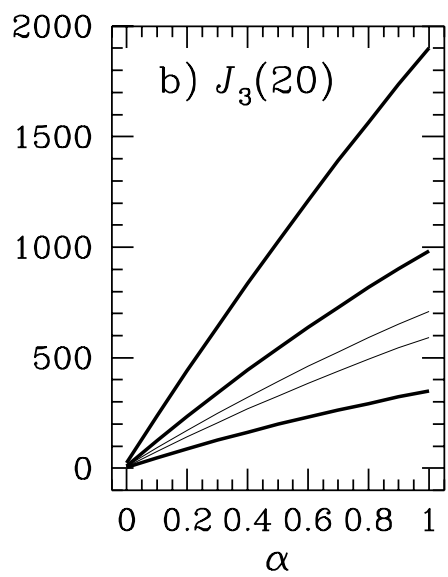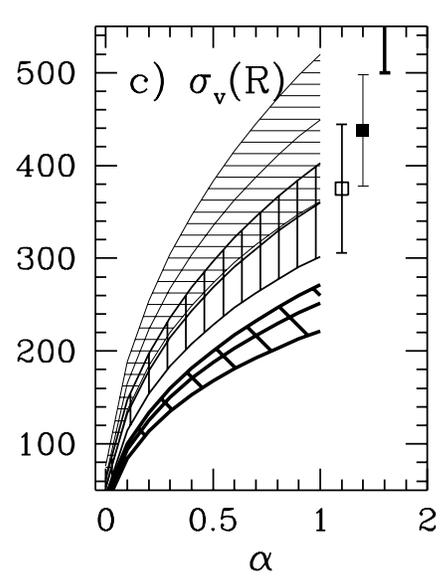

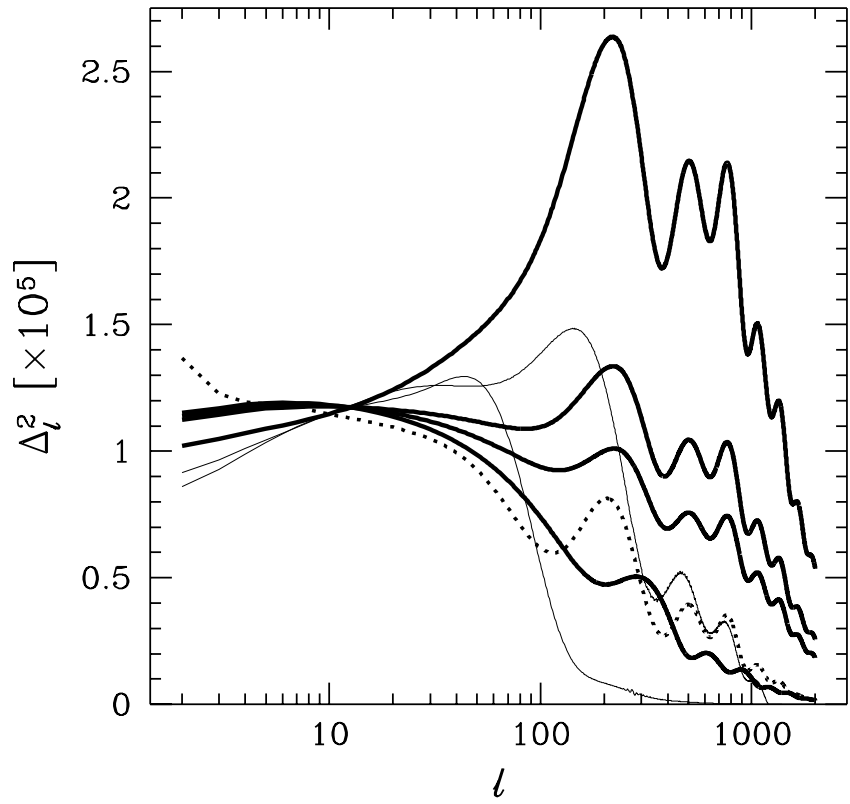

# FLAT DARK MATTER DOMINATED MODELS WITH HYBRID ADIABATIC PLUS ISOCURVATURE INITIAL CONDITIONS.

Radosław Stompor[1]

Copernicus Astronomical Center, Bartycka 18, 00-716 Warszawa, Poland,

E-mail: radek@camk.edu.pl

Anthony J. Banday

Hughes STX Corporation, 4400 Forbes Bldg, Lanham, MD 20706,

E-mail: banday@cmbr.gsfc.nasa.gov

and

Krzysztof M. Górski

Universities Space Research Association, NASA/GSFC, Laboratory for Astronomy and

Solar Physics, Code 685, Greenbelt, MD 20771

and Warsaw University Observatory, Aleje Ujazdowskie 4, 00-478 Warszawa, Poland,

E-mail: gorski@stars.gsfc.nasa.gov

---

[1]Present address : Department of Physics, University of Oxford, Keble Road, Oxford OX1 3RH, UK; electronic mail: rs@astro.ox.ac.uk

# ABSTRACT

We investigate the consequences of flat, dark-matter dominated cosmogonies with hybrid isocurvature and adiabatic initial perturbations and with Harrison-Zel'dovich primordial spectrum normalised to the $COBE$-DMR two-year measurements. We show that whilst the $COBE$-DMR data alone shows no preference for a specific admixture of these modes, acceptable combinations are strongly constrained by other observational data. Nevertheless, in some cases a suitable mixture of these modes still may be used in an attempt to avoid some of the observed problems of purely adiabatic models.

Specifically, we consider critical density, cold dark matter (CDM) and mixed dark matter (MDM) models. In the latter case we show that an isocurvature contribution of order $\sim 50 - 60\%$ (when expressed in terms of the contribution to the present day quadrupole) allows for a reasonably good fit to existing observational data. Low total density CDM-dominated models with any significant admixture of isocurvature mode are excluded, whilst CDM models with critical total density and hybrid initial conditions do not provide a significantly better fit to the data than purely adiabatic scalar models, and then only if the possibility of high amplitude pair-wise velocities on a $1h^{-1}$Mpc scale is considered.

*Subject headings:* cosmic microwave background — cosmology: observations — large-scale structure of the universe

## 1. Introduction

A normalisation scheme based on the *COBE* -DMR detection of cosmic microwave background (CMB) anisotropy results in serious problems for nearly all conventional models of structure formation in the Universe. The major difficulties of dark matter dominated, flat cosmogonies arise from the high normalisation amplitude required by the *COBE* -DMR detection of temperature anisotropies on large angular scales (Smoot et al. 1992, Bennett et al. 1994, Górski et al. 1994). For CDM models with a Harrison-Zel'dovich initial power spectrum this leads to excessive power on small-scales (Stompor, Górski & Banday 1995, hereafter SGB; Bunn, Scott & White 1995) as compared to the observational data. Such small-scale power can be suppressed by the addition of a hot dark matter admixture (hereafter MDM models) or non-zero cosmological constant. However, even the most plausible MDM models (Primack et al. 1995, Pogosyan & Starobinsky 1993, Ma & Bertschinger 1994) require a normalisation amplitude smaller than that inferred from *COBE* -DMR, while consideration of spatially flat models with a non-zero cosmological constant (CDM–$\Lambda$ ) yields serious constraints in their allowed parameter space (SGB). Some of these conclusions may be relaxed when a tilt in the primordial spectrum is invoked for the models –see Ostriker & Steinhardt 1995, White et al. 1995; section 6 for discussion. More exotic models have also been considered, none of which were found to be very attractive (e.g. Peebles 1994, Hu, Bunn, & Sugiyama 1995).

An attractive way to bypass the problem is to invoke a non-zero contribution from tensor perturbations in the early universe. Such modes contribute only to the large scale radiation anisotropy, thus allowing the amplitude of the matter perturbations to be smaller (Davis et al. 1992).

An alternative physically plausible mechanism which may reconcile both the amplitude of CMB anisotropies and matter perturbations invokes isocurvature initial conditions.

Unlike those models with adiabatic initial perturbations (scalar or scalar plus tensor), in which the large angular scale temperature anisotropies are generated by the Sachs-Wolfe (Sachs & Wolfe 1969; hereafter SW) effect, here the temperature fluctuations are mainly produced by photon overdensities on the last scattering surface – the isocurvature effect. Models of this type were previously considered either in the context of inflation (Efstathiou & Bond 1986, Sugiyama, Sasaki, & Tomita 1989) or topological defects (e.g. Brandenberger, Perivolaropoulos, & Stebbins 1990).

In this paper, we discuss dark matter dominated, critical density (CDM and MDM) models with hybrid (adiabatic plus isocurvature) Gaussian, scalar, initial conditions and a Harrison-Zel'dovich power spectrum, which are normalised by the *COBE* -DMR 2 year results. The relative amplitude of the modes is treated as a free parameter which we attempt to constrain using observations of the matter distribution in the universe.

## 2. Isocurvature versus adiabatic modes

The isocurvature character of the initial conditions means that the total density is almost unperturbed (i.e. the amplitude of the total density perturbations is much smaller than that of components) in the early moments of evolution. In the scenario we choose to focus on, the CDM overdensities are balanced almost perfectly by the radiation (i.e. photons and massless neutrinos) and baryon perturbations (e.g. photon-to-dark-matter entropy perturbations). This is a natural outcome of standard inflation in an axion-dominated universe (Steinhardt & Turner 1983; see Peebles 1994 for an analysis of the analogous isocurvature modes in the low total density, baryon-dominated case, and Hu & Sugiyama (1994) for a discussion of the complementary situation in which baryonic perturbations are balanced by those in the radiation). Compensation of the perturbations in different components remains true even at recombination only for those perturbations much larger then than the horizon. In particular, despite continuous pre-recombination growth (with growth rates $\propto a^3$ in the radiation epoch and $\propto a$ during matter domination), at recombination the amplitude of the superhorizon total density perturbations is much smaller than that of the photon perturbations (which grow at a rate $\propto a$ during radiation domination and then retain constant amplitude until horizon crossing). Consequently, the large angular scale CMB anisotropies in isocurvature models are dominated by the photon overdensities present on the last scattering surface (Efstathiou & Bond 1986) and not the SW effect which contributes only $\sim 16\%$ for $k \to 0$. The large scale temperature anisotropies expressed in terms of standard multipole coefficients are shown in Fig. 1a. The low-$\ell$ end of the radiation power spectra falls approximately as $[\ell(\ell+1)]^{-1}$ for both the adiabatic and isocurvature modes. This is because of the same scale dependence for both the SW and isocurvature effects ($\propto 1/k^3$) in the Harrison-Zel'dovich case. For $\ell \sim 30$, the slope flattens in the adiabatic case mainly due to the contribution of Doppler and adiabatic effects, which are much smaller on these scales for isocurvature modes. However, over the range of multipoles which $COBE$-DMR is sensitive to ($\ell \lesssim 30$), the radiation power spectra are very similar.

The superhorizon dark matter perturbations retain a constant amplitude during the radiation dominated epoch, but are damped during matter domination. After horizon crossing they rapidly fall into the potential wells generated by the total (matter plus radiation) overdensities. In a similar way, all sub-horizon baryon perturbations tend to fill potential wells soon after decoupling, while those on superhorizon scales retain their amplitude during matter domination. Therefore, the present-day matter power spectra (both dark and baryonic matter) on scales smaller than the current horizon resemble that of the total density at the present (Fig. 1). The latter is proportional to the total density power spectrum at recombination, given the scale independent post-recombination growth rate of the total density perturbations ($\propto a$ as in the adiabatic case). Hence, because of the much smaller amplitude of the total density perturbations at recombination (as a direct consequence of the applied normalisation), the present day matter power spectrum amplitude is smaller (by a factor of $\sim 36$) in the isocurvature case than for the adiabatic one. In addition, the isocurvature matter transfer function is more strongly damped on small-scales relative to large-scales than in the adiabatic case. Therefore the expected amplitude of the rms mass fluctuations in an $8h^{-1}$ Mpc sphere (denoted $(\sigma_8)_{mass}$) is roughly an order of magnitude lower than for adiabatic models. Since the latter $COBE$-DMR normalised models predict $(\sigma_8)_{mass} \lesssim 2.4$ (SGB) for a Hubble constant $h \lesssim 0.8$ (where $H_0 = 100h\,\mathrm{kms^{-1}Mpc^{-1}}$), pure isocurvature CDM models (for $0.3 \lesssim h \lesssim 0.8$) are equally unacceptable for notably underpredicting $(\sigma_8)_{mass}$ as adiabatic models (with $h \gtrsim 0.4$) are for overestimating its value. Also, in the isocurvature case the predicted bulk flows fall far below those deduced from POTENT (Dekel 1994), while the adiabatic models easily account for the measured value for almost all models of interest (SGB).

*Technical aside:* As a result of the very nature of the isocurvature perturbations, considered in this paper, at the present epoch the total density, baryon and dark matter matter power spectra are significantly different from each other on scales much larger than the actual horizon (but are identical on any sub-horizon scale). The Harrison-Zel'dovich character of the primordial total density power spectrum is preserved on the superhorizon scales, however it corresponds to the $\propto k^{-3}$ large scale behavior of the dark matter and baryon power spectra (with different coefficients). Therefore, the predicted $(\sigma_8)_{mass}$, in principle, are different for different components. Moreover, for the baryonic and dark matter power spectra, to avoid an apparent logarithmic divergence in the $\sigma_8$-integral, a long wavelength (i.e. small $k$) cut-off to the integral over $k$ has to be introduced. The predictions are, however, fairly insensitive to this value, increasing less than 1% for cut-off values ranging from the size of the present horizon to a value four order of magnitudes larger. Hence, the $(\sigma_8)_{mass}$ predictions are also independent of the choice of the matter component, as long as the superhorizon cut-off is assumed. Note that the quadrupole (as well as higher-$\ell$ multipoles) is always convergent and dominated by perturbations at subhorizon scales at the present. This holds also for bulk flows, which are determined by the total density distributions.

## 3. Hybrid modes

Although it is clear that invoking solely an isocurvature initial mode fails to reproduce the observational data succcessfully, an admixture of both isocurvature and adiabatic perturbations – by suppressing the small scale matter perturbations whilst retaining a reasonable match to the large-scale matter amplitude – could more favorably be compared to the observational constraints. Such a mixture could be generated during inflation (Steinhardt & Turner 1983), though it would require some fine-tuning of microphysical parameters. Hereafter, motivated by the inflationary model[2] (Axenides, Brandenberger, & Turner 1983, Steinhardt & Turner 1983), we assume the statistical independence of the amplitudes of both modes with the total density primordial power spectrum taken to be Harrison-Zel'dovich in shape. The mixed spectra are just a linear combination of the adiabatic and isocurvature spectra (Fig. 1b). Defining a mixing parameter $\alpha \in [0,1]$ with zero corresponding to a purely isocurvature mode, and unity to a completely adiabatic one,

---

[2]Note that a superposition of different perturbation spectra can also be achieved in adiabatic plus topological defect cosmologies (e.g. Pen, Spergel & Turok 1994) for which many of our comments would apply equally well. We are grateful to the anonymous referee for pointing this out to us.

we express the hybrid radiation and matter power spectra respectively as,

$$\begin{aligned} c_{(h)\ell} &= \frac{4\pi}{5} Q^2_{rms-PS}(\alpha)(\alpha \hat{c}_{(a)\ell} + (1-\alpha)\hat{c}_{(i)\ell}), \\ P_{(h)}(k) &= \frac{4\pi}{5} Q^2_{rms-PS}(\alpha)(\alpha \hat{P}_{(a)}(k) + (1-\alpha)\hat{P}_{(i)}(k)). \end{aligned} \quad (1)$$

The subscripts $a$, $i$ and $h$ denote adiabatic, isocurvature and hybrid mixture respectively. A hat over any quantity means that it was normalised in such a way as to make the radiation anisotropy rms quadrupole amplitude unity ($\hat{c}_2 = 1$). Quantities without the hat are normalised to the $COBE$-DMR 2 year data. Here, $Q^2_{rms-PS}$ ($= (5/4\pi)c_2 T_0^2$) is a straightforward generalisation of the normalisation convention introduced by Smoot et al. (1992) for pure power law models. The mode admixture is then expressed in terms of contributions to the present day quadrupole according to eqn. (1). The ratio of the primordial power spectra for the two modes depends on the model under consideration. For CDM, it can be expressed as $P_a(k,a_i)/P_i(k,a_i) \simeq 1.7 \times 10^7 (1-\alpha)\Omega_0^2 h^4/\alpha a_i^2$, which is the ratio of the adiabatic and isocurvature total density power spectra at an epoch $a_i$, much earlier than that of the matter-radiation equality (at the present, $a_0 = 1$), for any perturbations larger then than the horizon. Hence, the adiabatic total density fluctuations dominate the isocurvature ones at sufficiently early epochs.

## 4. Normalisation

We have applied the technique developed by Górski (1994), based on the computation of orthonormal basis functions for incomplete sky coverage and the subsequent expansion of the $COBE$-DMR two year sky maps (Bennett et al. 1994) in terms of these functions, to normalise all the models discussed in this paper. The spectra are fitted both with and without the quadrupole, which is the spectral mode most likely to be perturbed by faint high latitude galactic emission or (albeit small) residual systematic effects, but find that the difference in the estimates is always significantly smaller than statistical uncertainties.

The analysis is restricted to the ecliptic coordinate data since the systematic shift between normalisations inferred from either the galactic or ecliptic sky maps is also smaller than the statistical errors (see SGB for a complete discussion).

Since the radiation power spectra for both the adiabatic and isocurvature modes are similar over the range $\ell \lesssim 30$, we find that likelihood fits of the hybrid model to the data are only weakly dependent on the $\alpha$ parameter. The DMR data alone do not have sufficient power to rule out any particular admixture of adiabatic and isocurvature modes (Fig. 2). Although formally the likelihood fit to the data with the quadrupole excluded prefers an equal admixture of both modes, the result is not statistically significant.

The weak dependence of the $Q_{rms-PS}$ normalisation amplitude on the values of $\alpha$ and the Hubble constant can be conveniently summarised, to within a few per cent, by the relation,

$$\frac{Q_{rms-PS}(\alpha)}{Q_{rms-PS}(\alpha=1)} \simeq 1 + \beta(h)(1-\alpha) \geq 1, \qquad (2)$$

where $\beta(h)$ is approximated by the quadratic form $\beta(h) \simeq 0.13h^2 - 0.28h + 0.21$. This normalisation scheme is equally applicable to fits both with and without the quadrupole. The gentle increase of normalisation amplitude with growing isocurvature admixture is due to the slightly flatter shape of the low-$\ell$ end of the isocurvature radiation power spectrum. The errors on the fitted amplitudes are expected to be $\sim$ 10% ($1\sigma$ statistical error plus possible systematic effects) analogous to the pure adiabatic case (i.e. $\alpha = 1$). Here, SGB found that the value of $Q_{rms-PS}$ depended very weakly on cosmological parameters, and could be encompassed by the model-independent value $Q_{rms-PS}(\alpha = 1) = 20 \pm 2.2\mu$K.

## 5. Matter statistics

We consider a standard set of large-scale structure statistics computed within the framework of linear theory, namely $(\sigma_8)_{mass}$, $J_3$ and bulk flows variance $(\sigma_v^2)$ on POTENT scales. These are defined in linear theory as follows:

$$(\sigma_8^2)_{mass} \equiv \int_0^\infty dk\, W_{TH}(kR)P(k)k^2, \qquad R = 8h^{-1}\mathrm{Mpc},$$

$$J_3(R) \equiv (R^3/2\pi^2)\int_0^\infty dk\, k^2 W_{TH}(kR)P(k), \qquad R = 20h^{-1}\mathrm{Mpc}.$$

$$\sigma_v^2(R) \equiv (H_0^2/2\pi^2)\int_0^\infty dk\, W_{TH}(kR)e^{-kr_s^2}P(k), \quad r_s = 12h^{-1}\mathrm{Mpc},$$

Hereafter, $W_{TH}(x) = (3j_1(x)/x)^2$, and $P(k)$ stands for a total density power spectrum (see technical comment in paragraph 2).

It is easy to show that values of these statistics for the various hybrid models can be computed from the equivalent values determined for the pure isocurvature and adiabatic cases. For example,

$$\sigma_8(\alpha) = Q_{rms-PS}(\alpha)\left[\frac{\alpha\sigma_8^2(\alpha=1)}{Q_{rms-PS}^2(\alpha=1)} + \frac{(1-\alpha)\sigma_8^2(\alpha=0)}{Q_{rms-PS}^2(\alpha=0)}\right]^{1/2}, \qquad (3)$$

Analogous relations hold for other statistics.

Figure 3 shows the matter statistics as a function of the parameter $\alpha$. By decreasing $\alpha$ from unity to zero (i.e. a continuous transition from purely adiabatic to purely isocurvature perturbations) one reduces the amplitude of the matter power spectrum over the entire range of wavenumbers $k$, not only suppressing the small-scale power as desired, but also that on intermediate and large scales. The balance that one attempts to achieve by invoking mixed initial perturbation modes in critical density dark matter models is then to decrease the small-scale power so as, for example, not to overproduce pair-wise velocities, while at the same time retaining sufficient power on large scales to account for the bulk flow data. Below we discuss in more detail two interesting specific cases.

4.1 Hybrid CDM models

Here, we consider critical density models with the majority of the mass in the universe in the form of cold dark matter, a Hubble constant in the range $0.3 \lesssim h \lesssim 0.8$ and a baryon density $(\Omega_b = \rho_b/\rho_{crit})$ $\Omega_b = 0.0125h^{-2}$, to agree with recent constraints from nucleosynthesis (Reeves 1994).

N-body simulations are required to rigorously interpret small-scale constraints (since non-linear evolutionary effects will become important). Fortunately, it appears that the isocurvature mode admixture predominantly influences the amplitude of the spectrum, rather than its shape, so useful conclusions can be inferred from results derived for the standard CDM scenario.

It is well known that to account for the amplitude of the observed pair-wise galaxy velocity dispersion (Davis & Peebles 1983) a large bias factor $b = (\sigma_8)_{gal}/(\sigma_8)_{mass} \gtrsim 2$ (and therefore $(\sigma_8)_{mass} \lesssim 0.5$) is required. CDM models with $h \gtrsim 0.5$ can only match this requirement if the contribution of the isocurvature mode is such that $\sqrt{1-\alpha} \gtrsim 0.95$. However, the corresponding large scale flows would be significantly underproduced, e.g. $\sigma_v(R = 60\,h^{-1}\mathrm{Mpc}) \lesssim 150\,\mathrm{km\,s^{-1}}$. Thus only models with $h \sim 0.3$ (and practically pure adiabatic initial conditions) would survive. This is simply the small vs. large scale structure incompatibility noted even before the *COBE* era.

Marzke et al. 1995, have recently re-analysed the pair-wise velocities and found that acceptable values may be as high as $500 - 600\,\mathrm{km\,s^{-1}}$ (as suggested by both Mo et al. 1994, and Zurek et al. 1994). $(\sigma_8)_{mass}$ can thus be increased to $0.6 - 0.7$ which would marginally satisfy the POTENT constraints (Zurek et al. 1994, Seljak & Bertschinger 1994). The isocurvature mode contribution, $\sqrt{1-\alpha}$, might optimistically be reduced to 0.9 even for models with $h$ as high as 0.5, but even then the large-scale galaxy clustering or cluster correlations (see e.g. Efstathiou 1995 and references therein) remain problematic.

The most stringent upper bound on the isocurvature mode contribution arises from the attempt to reproduce the observed large-scale flows (requiring $\alpha \gtrsim 0.2$ to match the POTENT velocities, whilst the Lauer-Postman datum – Lauer & Postman, 1994 – remains problematic). Such a constraint has immediate consequences for the possible isocurvature contribution to low density, spatially flat (i.e. cosmological constant dominated) models, practically excluding any appreciable admixture.

*4.2 Hybrid MDM models*

For $\alpha \gtrsim 0.2$ the matter characteristics computed for hybrid models are dominated by the adiabatic contribution and the formula (3) can be replaced to sufficient accuracy by the relation:

$$\sigma_8(\alpha) \simeq \sqrt{\alpha}\,\sigma_8(\alpha = 1)(1 + (1-\alpha)\beta(h)), \quad \alpha \gtrsim 0.2. \qquad (4)$$

The adiabatic mode is thus sufficient to make reasonable predictions even in hybrid cases.

Since the small-scale power enhacement is less pronounced in MDM models, the contribution of the isocurvature mode does not need to be as dominant as in the CDM situation. For MDM, a $\sim 20-30\%$ suppression of small-scale power is sufficient, and our approximate formula (4) is applicable.

Table 1 of SGB lists the important large-scale linear statistics for a number of adiabatic MDM models. Two of the more interesting choices of cosmological parameters made recently are: (i) $h = 0.5$, $\Omega_b \simeq 0.05$ with one massive neutrino family $m_\nu \simeq 7\text{eV}$ contributing to the total density $\Omega_\nu = 0.3$, and cold dark matter with density $\Omega_{CDM} = 1 - \Omega_b - \Omega_\nu = 0.65$ (Pogosyan & Starobinsky 1993); and (ii) $h = 0.5$, $\Omega_b \simeq 0.075$, with two families of massive neutrinos with $m_\nu \simeq 2.4\text{eV}$ each, $\Omega_\nu = 0.3$ and $\Omega_{CDM} = 0.625$ (Primack et al. 1995).

Model (i) normalised to the *COBE*-DMR 2 year maps (SGB) yields a linear theory estimate of $(\sigma_8)_{mass} \simeq 0.85$. Pogosyan & Starobinsky claim that a more desirable value to reproduce the observational data is close to 0.6 (corresponding to $Q_{rms-PS} \simeq 15\mu\text{K}$), while models with a higher $(\sigma_8)_{mass}$ are somewhat disfavoured. An isocurvature mode admixture ($\sqrt{1-\alpha} \sim 0.55 - 0.6$) would both suppress the excess $(\sigma_8)_{mass}$ amplitude and reconcile the model with the inferred $COBE$-DMR normalisation. Subsequent analyses (e.g. Nusser & Silk 1993, Ma & Bertschinger 1994) have questioned, however, the ability of models with such a high HDM contribution (and one family of massive neutrinos) to account for the abundances of high redshift objects. Since lowering the HDM abundance with a normalization fixed by the COBE-$DMR$ 2 year data increases appreciably the predicted $(\sigma_8)_{mass}$, such models would require a still larger isocurvature admixture. While this could solve the problem on $8h^{-1}$Mpc scales, it would decrease the bulk velocities in a manner similar to that discussed in the CDM case. With hybrid initial conditions, even MDM models with a small HDM contribution (e.g. as provided by very light neutrinos $m_\nu \sim 2/N_\nu$ eV, where $N_\nu$ –number of massive neutrino families) can be reconciled with the observations, resolving most of the problems of a purely CDM model.

Model (ii) normalised to $COBE$ predicts a value for $(\sigma_8)_{mass}$ of 0.82, significantly higher than the 'successful' value of 0.6 (see Primack et al. 1995). The discrepancy can again be addressed by considering hybrid initial conditions with $\alpha \sim 0.7 - 0.75$.

## 6. Discussion

We have demonstrated that the adiabatic or isocurvature nature of the density perturbations in the early universe can not be resolved on the basis of the $COBE$ -DMR 2 year data alone. Rather, we are able to treat the relative amplitude of both modes as a free parameter which can be adjusted to improve the fits of dark matter dominated flat cosmogonies to the observational large-scale structure data. Such an improvement is largely a consequence of the isocurvature effect generating the large angular scale CMB anisotropy in isocurvature models, whilst the small-scale matter power is simultaneously suppressed relative to adiabatic models.

For CDM models, even in the most optimistic situation, a large isocurvature admixture ($\sqrt{1-\alpha} \sim 0.9 - 0.95$) is required to resolve the velocity-dispersion problem, but then can not simultaneously account for the large scale matter distribution. The shape of the CDM matter power spectrum, even incorporating hybrid initial conditions, is not a good match to the observed galaxy power spectrum, as was known during the pre-*COBE* epoch. The situation can be improved for purely adiabatic models by the introduction of a hot dark matter component (Pogosyan & Starobinsky 1993, Primack et al. 1995). Nevertheless, the predicted amplitude of the small scale perturbations remains high, and an admixture of the isocurvature mode, $\sqrt{1-\alpha} \sim 0.55$, may still be necessary to reconcile the linear theory predictions for MDM models with the observational data. MDM models with a low contribution of HDM (as required by the observed abundances of high redshift objects) can be also accommodated within the proposed scenario and satisfy observational constraints as long as $\Omega_{HDM} \gtrsim 0.1$.

More precise and direct bounds on the size of the isocurvature contribution may be obtained by precise measurements of CMB temperature fluctuations on angular scales of a degree and below (i.e. $\ell \gtrsim 100$). Whilst the suppressed height of the radiation spectrum ('Doppler') peaks with respect to the SW plateau does not uniquely testify to the existence of an isocurvature contribution (Fig. 4), the relative heights of the first two peaks and the SW plateau, together with their angular positions, may be uniquely indicative of hybrid initial conditions.

It might appear that the introduction of the proposed isocurvature mode produces qualitatively similar results to the addition of a gravity wave contribution. Nevertheless, there are several key differences between the scenarios. A significant one is that the contribution of gravity waves depends on the spectral index of the primordial perturbations, vanishing exactly in the Harrison-Zel'dovich ($n = 1$) case (Davis et al. 1992). Thus a tilted spectrum is a pre-requisite for a significant gravity wave component to the CMB temperature anisotropy. Conversely, the isocurvature mode admixture influences only the smallest scales, and in the range of $\alpha$ of interest, the changes are quite modest.

The possibility of tilted CDM cosmologies (both with and without a tensor mode contribution) was recently considered by White et al. 1995, who found that a low Hubble constant ($h \lesssim 0.5$) model with $n = 0.8 - 0.9$ could provide a satisfactory fit to tha data, but only if a small bias ($b \sim 1.2$) suffices to account for pair-wise velocities, which remains to be seen. The required tilt may be even lower for MDM models, but the additional small-scale power suppression can cause an underproduction of high redshift objects – a test only marginally passed by the $n = 1$ MDM models (Ma & Bertschinger 1994).

From the point of view of the inflationary scenario, tilted spectra (with or without gravity waves) seem to be a more natural solution to matching large and small-scale structure constraints, while an appreciable admixture of both adiabatic and isocurvature modes would seem to require some fine-tuning of fundamental physical parameters unrelated by any currently understood physical mechanisms (Steinhardt & Turner 1983). Nevertheless, we know of no physically compelling reason to definitively preclude the existence of such an admixture.

We acknowledge the efforts of those contributing to the *COBE*-DMR. This work was supported in part by the Office of Space Sciences of NASA Headquarters. RS was supported in part by the Polish Scientific Committee (KBN) grant no. 2P30401607. RS and KMG were supported in part by the National Science Foundation under Grant No. PHY94-07194.

**Figure captions:**

Figure 1. The radiation (a) and total density (b) present-day power spectra (solid lines) for cold dark matter dominated flat models with $h = 0.5$, $\Omega_b = 0.05$ and hybrid initial conditions. We define $\Delta_\ell \equiv \sqrt{\ell(2\ell+1)c_\ell/4\pi}$. In the left panel, the mode mixing ratio ($\alpha$) is zero (pure isocurvature mode) for the curve with the highest quadrupole ($c_2$) and unity for the lowest one (pure adiabatic mode). This trend is reversed in the right hand panel. In (b) the large scale $k^{-3}$–tails of the CDM (short dash) and baryon (long dash) power spectra are displayed. The arrow indicates the wavenumber corresponding to the size of the present-day horizon.

Figure 2. Likelihood fits of the hybrid models with $h = 0.5$ to the *COBE*-DMR 2 year data (in *ecliptic coordinates*) as a function of the $Q_{rms-PS}$ for different values of the mixing parameter $\alpha$: dashed lines quadrupole included, solid lines quadrupole excluded. Left to right the three curves for each linestyle correspond to a purely adiabatic model, an equal admixture of adiabatic and isocurvature modes, and a purely isocurvature model respectively.

Figure 3. The predicted matter statistics for models with hybrid initial conditions as a function of the mixing ratio $\alpha$. (a) & (b) The thick solid lines are for CDM dominated flat models with $h = 0.8, 0.5, 0.3$ and $\Omega_b = 0.0125 h^{-2}$ (from top to bottom), while thin solid lines correspond to MDM models with parameters $h = 0.5$, $\Omega_b = 0.05$, $\Omega_{HDM} = 0.3$ & $N_\nu = 1$ (top line) and $h = 0.5$, $\Omega_b = 0.075$, $\Omega_{HDM} = 0.2$ & $N_\nu = 2$ (bottom one). The dotted lines (which overlap the solid lines except for $\alpha \lesssim 0.2$) are approximations given by formula (4). (c) The predicted bulk velocities in flat CDM models. The three shaded regions correspond to values of a top hat filter (modified by an additional gaussian

smoothing on a scale of $12h^{-1}$Mpc in each case) with radius equal to $40h^{-1}$Mpc (upper shaded region), $60h^{-1}$Mpc (middle region), $100h^{-1}$Mpc (lower), and Hubble constant in the range $0.3 \leq h \leq 0.8$, with the upper edge of the region corresponding to $h = 0.8$, the lower one to 0.5, and the middle line to 0.5. The predictions for MDM models are very close to the CDM values for the same Hubble constant. Squares denote the POTENT result on scales $40h^{-1}$Mpc (solid) and $60h^{-1}$Mpc (open). The $1\sigma$ lower limit from the Lauer & Postman (1994) measurements is also shown.

Figure 4. The radiation matter spectra for flat CDM dominated models with $h = 0.5$, $\Omega_b = 0.05$ and mixing ratio $\alpha$ equal to 0 (the lowest thick line), 0.1, 0.2, and 1 (the highest thick line). This is a progression from purely isocurvature to completely adiabatic initial conditions. For comparison, we also show power spectra for variations on the flat CDM model (with the same background parameters). These include: CDM adiabatic models with non-standard thermal scenarios in which the residual free-electron number density after recombination equals 0.05 and 0.1 (the thin curves peaked at $\ell \sim 150$ and at $\sim 40$, respectively), and a tilted adiabatic CDM model with $n_{scalar} = 0.7$ and the tensor to scalar mode ratio given by $T/S = 7(1 - n_{scalar})$ (dotted line). $\Delta_\ell$ is defined as in Figure 1a.